\documentclass[fleqn,usenatbib]{mnras}

\usepackage[T1]{fontenc}
\usepackage{ulem}
\DeclareRobustCommand{\VAN}[3]{#2}
\let\VANthebibliography\thebibliography
\def\thebibliography{\DeclareRobustCommand{\VAN}[3]{##3}\VANthebibliography}

\usepackage{graphicx}	% Including figure files
\usepackage{amsmath}	% Advanced maths commands
\usepackage{amssymb}	% Extra maths symbols
\usepackage{newtxtext,newtxmath}
\usepackage{xcolor}
\usepackage{float}
\usepackage{tabularx}
\usepackage{color,soul}

\usepackage{subcaption}
\usepackage{lipsum}

\title[One-point and Environments]{One-point Statistics in various cosmic environments in the presence of massive neutrinos}

\author[Khoshtinat et al.]{
Mohadese Khoshtinat$^{1}$, 
Hossein Hatamnia$^{1}$, 
Shant Baghram$^{1, 2}$ \thanks{baghram@sharif.edu}
\\ \\
$^{1}$
Department of Physics, Sharif University of
Technology, P.~O.~Box 11155-9161, Tehran, Iran\\
$^{2}$
Research Center for High Energy Physics, Department of Physics, Sharif University of Technology, P.O.Box 11155-9161, Tehran, Iran\\
}

\date{Accepted XXX. Received YYY; in original form ZZZ}

\pubyear{2024}

\begin{document}
\label{firstpage}
\pagerange{\pageref{firstpage}--\pageref{lastpage}}
\maketitle
\begin{abstract}
Studying the structures (halos and galaxies) within the cosmic environments (void, sheet, filament, and node) where they reside is an ongoing attempt in cosmological studies. The link between the 
properties of structures and the cosmic environments may help to unravel the nature of the dark sector of the Universe. In this paper, we study the cosmic web environments from the spatial pattern perspective in the context of $ \Lambda $CDM and $ \nu \Lambda $CDM as an example of an extension to the vanilla model. To do this, we use the T-web classification method and classify the cosmic environments for the catalogues from the gevolution N-body simulations for $ \Lambda $CDM and $ \nu \Lambda $CDM cosmology. Then, we compute the first nearest neighbour cumulative distribution function,  spherical contact cumulative distribution function, and $ J$-function for every cosmic environment. In the context of the standard model, the results indicate that these functions can differentiate the various cosmic environments. In association with distinguishing between extensions of the standard model of cosmologies, these functions within the cosmic environment seem beneficial as a complementary probe.
\end{abstract}


%%%%%%%%%%%%%%%%%%%%%%%%%%%%%%%%%%%%%%%%%%%%%%%%%%

%%%%%%%%%%%%%%%%% BODY OF PAPER %%%%%%%%%%%%%%%%%%

\section{Introduction}
The $\Lambda$CDM model of cosmology successfully predicted almost all results obtained from the cosmic microwave background (CMB) and the large-scale structure (LSS).
But as the data from ongoing surveys  \citep{Planck:2018vyg, SDSS:2023tbz, DESI:2023ytc, Euclid:2024vss} become accurate and rich on the finer scales, the observed quantities deviate from the predicted values based on the $\Lambda$CDM model. Some of the deviations become serious and challenge the $\Lambda$CDM model, e.g., the Hubble tension \citep{di2021realm}, the $\sigma_8$ tension \citep{Joudaki:2017zdt} together with the discrepancy on small scales \citep{Perivolaropoulos:2021jda}.
{{These challenges, along with the principal questions about the physical nature of the dark sector of the Universe, created a rich literature around the extension of the standard model of cosmology. The decided extension is to consider the massive neutrinos as a component of the dark matter--$\nu \Lambda$CDM since the neutrino oscillation phenomena establish that the total neutrino mass is nonzero \citep{KATRIN:2021uub}.  One of the objectives of the next generation 
cosmological experiments is to measure the total neutrino mass \citep{Lesgourgues:2006nd}. Some other speculative extensions are the dynamical dark energy model \citep{Peebles:2002gy, Tabatabaei:2023qxw, Amendola:2015ksp}--$\omega \Lambda$CDM, interacting dark matter and energy \citep{Amendola:1999er}, and the deviation from the Gaussian initial condition \citep{Bartolo:2004if}.}}
Most studies so far apply these models to the datasets using the traditional tools used by cosmologists, i.e., the 2-point correlation functions in real space or its Fourier counterpart, the power spectrum. The 2-point correlation function is the best choice for investigating the cosmological dataset up to the scales in which the linear perturbation theory still holds \citep{Bernardeau:2001qr}. 
Beyond that, due to highly nonlinear gravitational evolution and heavy mode coupling, the information leaks to the higher-order correlation function, and the 2-point correlation function fails to capture all the information hidden in the dataset. The leaked information at these scales is crucial as the deviation from the $\Lambda$CDM model occurs mostly at these scales. To extract this information, one can follow two widely used strategies. 
One is to equip the 2-point correlation with extra details accounting indirectly for the higher-order information. Some examples are the marked power spectrum for halos or galaxies \citep{stoyan1984correlations, Massara:2020pli}, the redshift space studies \citep{Kaiser:1987qv}, and computing the 2-point function in different cosmic web environments and components---void, sheet, filament, and node---in the coordinate and the redshift spaces \citep{Bonnaire:2021sie, Bonnaire:2022ocm}. The latter highlights the potential of environmental studies on constraining the cosmological parameters and breaking the degeneracies between different extensions of the $ \Lambda$CDM model. 
\cite{Zeldovich:1969sb}, and many others, e.g., \cite{doroshkevich1978statistical},  noted that the formation of these structures is rooted in the collapse of anisotropic fluctuations in a specific environment via gravitational instabilities in an expanding universe.
 \\
 {{The second solution to get the most out of data is to find other interpretable statistics that, in principle, contain all the information of n-point correlations but computed for one point, e.g.,  the density-split statistics that compute the probability distribution function (PDF) of the late-time matter density field based on the count of tracer galaxies \citep{DES:2017eav}, the PDF of the matter density field smoothed at some radius \citep{Uhlemann:2019gni, Uhlemann:2022znd}. Moreover, \cite{White:1979kp} used the void probability function to study the clustering of the tracers, \cite{Banerjee:2020umh, Banerjee:2021cmi, Yuan:2023llf} introduced the k-nearest neighbours cumulative distribution functions (CDF) or peaked-CDF to study the properties of the non-Gaussian clustering at the small scale.}}
 \\
\cite{Fard:2021qaa} and \cite{Khoshtinat:2023zck} used the letter functions to study the clustering of the dark matter halos in the context of the $ \Lambda$CDM and the $ \nu \Lambda$CDM, respectively and \cite{Kousha:2023kog} for studying the distribution of LSS in Fuzzy dark matter models. The letter functions---$ G(r)$ or the nearest neighbour CDF,   and $ F(r)$ or the spherical contact CDF--- are statistics used to investigate the clustering aspect of the distribution of points in the space. \\
In this paper, we study the behaviour of the letter functions in various cosmic web environments identified by the T-web classification method  \citep{Hahn:2007ui,  Forero-Romero:2008svv} in the context of the $ \Lambda$CDM and then use the $ \nu \Lambda$CDM as an example of the extended model of cosmology. In Section \ref{Sec2}, we summarize the theoretical ingredients used in this work. Section \ref{Sec3} describes the simulations and the algorithm used to identify the cosmic environments. Section \ref{Sec4} represents the results of the analysis. Finally, in Section \ref{Sec5}, we conclude the discussion and point to the future direction.

\section{Theoretical Background}\label{Sec2}
The following section provides essential theoretical aspects required for this work. The first subsection discusses the cosmic web, the second addresses the impact of the neutrino mass on cosmology, and the final one covers one-point statistics.

%%****************FIG 1**********************%%
\begin{figure}
\centering
\includegraphics[width=0.48\textwidth]{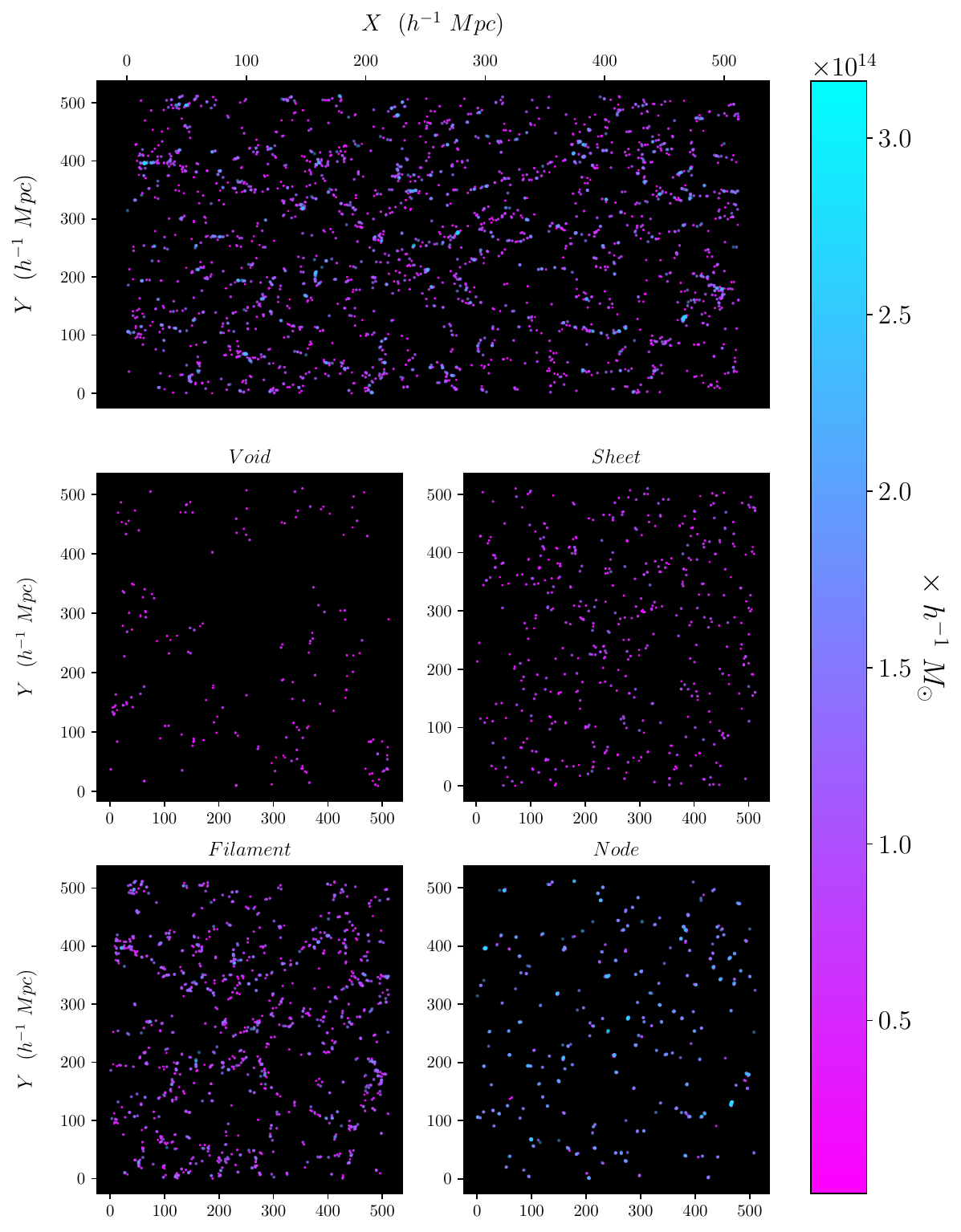}
\caption{A $ 5 \ h^{-1} \rm{Mpc}$ thick slice of one sub-box of the $\Lambda$CDM simulation at $ z=0$. The topmost panel shows all the halos in the slice. The other four illustrate the halos in various environments: voids, sheets, filaments, and nodes, respectively. The T-web formalism is used to identify the environments.} 
\label{fig:Env-Autumn}
\end{figure}
%%****************FIG 1**********************%%

\subsection{Cosmic Web Environment}\label{sec:CWE}
It was evident from early galaxy surveys to recent ones \citep{Huchra:1983wy,deLapparent:1985umo, SDSS:2002oin,SDSS:2008tqn} and also from the N-body simulations \citep{Springel:2005mi} that the distribution of matter in the Universe displays a filamentary pattern, known as the cosmic web. The theoretical description---Zeldovich approximation---of the formation of such a filamentary configuration goes back to the works of \cite{Zeldovich:1969sb, doroshkevich1978statistical, arnold1982large}. Zeldovich approximation predicts the formation of the cosmic web components due to the local level of tidal anisotropy for the linear growth of perturbation, i.e., in the Lagrangian description based on the series expansion of the equation of motion, test particles along their trajectories experience zero-order acceleration and first-order deformation determined by the tidal field, which ultimately leads to the formation of the voids, sheets, filaments, and nodes. This physical picture sets a mathematical guideline for classifying the cosmological environment based on the eigenvalues of a deformation tensor \citep{Libeskind:2017tun, Aycoberry:2023uyk} and a venue to count the number of cosmic structures \citep{Sheth:2003py,Colberg:2004nd,Fard:2018dwx}. Among various definitions for a deformation tensor, we use the T-web classification scheme extended by \cite{Forero-Romero:2008svv}. In this approach, they defined the tidal shear tensor $ T$ as follows: 
\begin{equation} \label{T-tensor}
\\
T_{ij} = \frac{\partial^2 \Phi}{\partial x_i \partial x_j},
\end{equation}
 where $ \Phi$ is the gravitational potential, $ i$ and $ j$ stands for the spatial coordinates.
The environments can then be classified using the following criteria for the eigenvalues of the $ T$ tensor:
\begin{itemize}
\item \textbf{Node:} \textit{Three} eigenvalues are above the threshold.
\item \textbf{Filament:}  \textit{Two} eigenvalues are above the threshold. 
\item \textbf{Sheet:} \textit{One} eigenvalue is above the threshold.
\item \textbf{Void:} \textit{No} eigenvalue is above the threshold.
\end{itemize}
The threshold is a free parameter of the process, obtained by complementary observations such as the number density of each element in the cosmic web or their volume fraction. We will come back to it in Section \ref{sec:CW-Classif}. 
\\
In the context of the standard model, analysis of the properties of the halos from the numerical simulation \citep{Hahn:2007ui, Wechsler:2005gb} suggests that these are not only dependent on the mass of the halo (as considered in the semi-analytical models like the excursion set theory \citep{Bond:1990iw, Zentner:2006vw}) but on the environments where the halo is formed and evolve during the cosmic time \citep{Sheth:2004vb,Parkavousi:2022llz}. At each epoch, the heaviest halos residue in the nodes having the broadest range of halo mass decades contrasted to the other environments, with the voids having the shortest mass interval. Additionally, \cite{Hahn:2007ui} observed that the typical mass-scale collapsing at each epoch, $ M_{col}(z)$, acts as a natural criterion for the halo mass. Halos that are affected the most by the characteristics of their surrounding environments are those with masses less than  $ M_{col}(z)$.  
For example, halos with a mass less than $ M_{col}(z)$ tend to align more with the preferred direction of their hosting filament or sheet, spin-wise or shape-wise. 
Another notable characteristic is that the voids have the youngest halos with the mass $ \lesssim M_{col}(z) $ as opposed to the other environments \citep{Hahn:2007ui}.
With these observations in mind, in this work, we study the environment dependence of one-point statistics.

\subsection{Massive Neutrino}
The neutrino oscillation phenomenon is an experimental fact that the total mass of the three neutrino species can not be considered zero. The latest release from the KATRIN \citep{KATRIN:2021uub} experiment suggests an upper bound for the total neutrino mass of about $ \Sigma \ m_{\nu} \lesssim0.8$ eV ($90 \% $ C. L.). Regarding the neutrinos, their mass puts them in the category of the light relic in cosmological studies. 
Light relic species leave the primordial thermal equilibrium as relativistic particles. As the Universe expands, these particles lose their energy and eventually become non-relativistic when their mass surpasses their kinetic energy. This procedure induces a length scale---$ \lambda_{fs}$, below which these particles stream freely because of their thermal velocities. For a single neutrino species $ m_{\nu}$ at any given redshift $ z$, this length scale is defined accordingly:
\begin{equation}
\\
k_{fs} (z) \approx 0.063 \, h \  {\rm Mpc}^{-1} \, \frac{m_{\nu}}{0.1 {\rm eV}} \ \frac{1}{(1+z)^2} \ \frac{H(z)}{H_0}\ , 
\end{equation} 
where $ H(z)$ and $H_0 $ are Hubble parameter and constant, respectively \citep{Agarwal:2010mt}.
The free-streaming of neutrinos below $ \lambda_{fs} \sim k_{fs}^{-1} $ smooths the fluctuation of the matter field, which translates to power suppression contrasted to the massless scenario. Beyond the free-streaming length scale, neutrinos behave like cold dark matter particles. At redshift zero, the neutrino energy density is equal to:
\begin{equation} \label{eq-neutrino_density}
\\
\omega_{\nu} = \frac{ \sum m_{\nu} }{ 94.13 \ \rm  eV} \ ,
\end{equation} 
with the neutrino mass fraction of $f_{\nu} = {\omega_{\nu}}/{\omega_m} $, where $\omega_m=\Omega_m h^2$ is the dark matter density normalized to the critical density.
\\
In addition to the suppression of the relative power spectrum ($\nu \Lambda $CDM to the standard model), predicted by both linear and non-linear perturbation theory, data from the N-body simulations with massive neutrinos show a turnover of the relative power spectrum around $ k \sim  1 \ h  \ {\rm Mpc}^{-1}$, also known as \textit{spoon-like} as opposed to predicted \textit{slide-like} behaviour. \cite{Hannestad:2020rzl} claims from the halo model \citep{Navarro:1995iw, Cooray:2002dia} perspective suppression of the relative power spectrum is due to the suppression of the two-halo clustering caused by neutrinos. Whereas the rise in the relative power spectrum at $  k \sim 1 \ h  \ {\rm Mpc}^{-1}$ is controlled by the one-halo term, where the virialization process eradicates the initial condition of perturbations deep inside a halo. 
Regarding the spatial distribution of halos, \cite{Khoshtinat:2023zck} observe for a randomly placed point in a universe with massive neutrinos, the neighbouring is empty than the $ \Lambda$CDM universe, and the degree of emptiness highly depends on the total neutrino mass. They also show that the immediate neighbour of halos in the presence of massive neutrinos happens at further distances as opposed to the nearest neighbour for the $ \Lambda$CDM universe. Concisely, the presence of massive neutrinos reduces the overall clustering in the LSS.
These observations motivate us to study the aforementioned effects in different cosmic environments.

\subsection{One-point Statistics}\label{1-pointstat}
The spatial distribution of the dark matter halos, formed by gravity, carries loads of information about the underlying dark matter field \citep{ravi1996distribution,  Bardeen:1985tr}. In the absence of gravity  and other kind of interactions, the pattern exhibited by the halos would be an example of the Poisson point process\footnote{ For instance, a set of $N$ points in a cubic box of side $L$ generated from a uniform distribution is an example of a Poisson process with the intensity of  $ \bar{n} = NL^{-3}$.}, i.e., complete spatial randomness. 
The main characteristic of a Poisson point process is the absence of clustering, which is not the case for the LSS in the Universe. 
 In the context of the spatial point analysis \citep{hand2008statistical}, the letter functions--- $ G(r)$, $ F(r)$, and $ J(r)$---are tools used to examine the clustering and its extent for a spatial distribution of points. The $ G$ and $ F$ are of the nature of cumulative distribution functions and are defined as follows:
\begin{itemize}
\item $ G(r)$: The CDF of the first nearest neighbour distance between the halos. 
\item $ F(r)$: The CDF of the distances from randomly generated points to the first neighbouring halos.
\end{itemize}
$ G(r)$, for a single halo, means the probability of having its first nearest neighbour on the surface of a sphere with the radius $ r$. 
Whereas for all halos in the catalogue, it indicates the proportion of the halos having their first nearest neighbour on a sphere of radius $ r_i \le r $ around them where $r_i$ stands for the sphere centered at the location of $i$-th halos. $ F(r)$ states the same but for the randomly generated points in the corresponding volume. $J-$function is defined:
\begin{equation} \label{Eq-Jfun}
\\
J(r) \equiv \frac{1 - G(r)}{1 - F(r)}.
\end{equation}
The value of the $ J$-function for complete spatial randomness is one, while in the presence of clustering, it takes values less than unity.

Aimed at studying the large-scale structure, one should address the information these statistics hold and their connection to n-point correlation functions. In the seminal work of \cite{White:1979kp}, the conditional correlation functions $ \Xi_i $ defined as,
\begin{align} \label{Eq-conditional corr}
\Xi_i(\textbf{r}_1,...,\textbf{r}_i;V) =& \sum_{j=0}^{\infty} \dfrac{(-n)^j}{j!} \\ \nonumber
& \int ...\int \xi_{i+j}(\textbf{r}_1,...,\textbf{r}_{i+j})\,dV_{i+1}...dV_{i+j},
\end{align} 
establish a way to link these statistics and the n-point correlations. In equation \ref{Eq-conditional corr}, $ \xi_i $ are the n-point correlation functions where by definition $\xi_0 = 0$ and $\xi_1 = 1$. And the integrals are over an empty volume of interest $V$, except for the points $ r_i $ and  $ \xi_i $.
\\
The void probability, $ P_0(V)$--the probability of finding an empty sphere of radius $ r$, is defined as $ P_0(V) = \exp\left[ \Xi_0(V(r))\right] $. By definition, the complement of $ P_0(V)$ is the $ F(r)$.  
\\
The probability for a halo to have its nearest neighbour on the surface of a volume $ V(r)$ is the complement of the absence of any halo in this volume, $ G(r) = 1 - \Xi_1(\textbf{r}_o; V(r)) \exp\left[ \Xi_0(V(r))\right] $.

These relations and equation \ref{Eq-conditional corr} show that, by definition, these statistics depend on the correlations of all orders, which makes them a suitable probe to study the LSS, especially at the non-linear scales.

\section{Simulations and Cosmic Web classification}\label{Sec3}
In the first subsection, we introduce the simulation suit used in this work. In the subsequent subsection, we review the details of the cosmic web classification procedure.

%%****************Table 1**********************%%
\begin{table}
\begin{center}
\begin{tabular}{ c | c | c | c | c | c }
$\omega_c$ & $\omega_b$ & $h$ & $n_s$ & $T_{\rm CMB}$ &  $ A_s $ \\ 
\hline \hline
$0.12038$ & $0.022032$ & $0.67556$ & $0.9619$ & $ 2.7255$ & $2.215 \times 10^{-9}$ \\ 
\end{tabular}
\caption{The cosmology parameters of the standard model where $\omega_c=\Omega_ch^2$, $\omega_b=\Omega_b h^2$ are baryonic and dark matter densities normalize to the critical density. $h=H_0 / 100$, $n_s$, $T_{\rm CMB}$, $ A_s$ are respectively the reduced Hubble parameter, spectral index, the CMB temperature, and the amplitude amplitude of scalar perturbations.}
\label{Table1}
\end{center}
\end{table}

%%****************Table 1**********************%%

\subsection{Simulations}
We use existing halo catalogues for the $\nu \Lambda$CDM models with a suitable $ \Lambda$CDM reference catalogue to study the environmental effects using the letter functions. Each catalogue is prepared from a snapshot of the gevolution \footnote{\url{https://github.com/gevolution-code}} N-body simulation \citep{Adamek:2016zes, Adamek:2015eda} by the ROCKSTAR halo finder \citep{Behroozi:2011ju}. The simulations follow $4096^3$ particles and grids in a cubic comoving box $L= 2048$ $h^{-1} {\rm Mpc}$ initialized at redshift $z=100$. The total neutrino mass in the $\nu \Lambda$CDM catalogues are $\sum m_{\nu} = 0.06, \ 0.2, \ 0.3$ eV. The cold dark matter density parameter for each of these catalogues is consistent with the $\omega_c \equiv \Omega_c h^2= 0.12038 - \omega_{\nu}$, so the total matter density ($\omega_{\nu}+\omega_{b}+\omega_{c}$) is untouched. 
{{Since gevolution relies on the weak-field expansion of general relativity, it can treat light relics such as neutrinos as particles and follow their non-linear evolution. In this simulation suit, neutrinos are simulated using ensembles with $ 1.7 \times 10^{11} $ tracers to construct an effective stress-energy tensor. This tensor is then used in the weak-field formulation of general relativity, which the gevolution simulation relies on. For more detailed information on the simulation, refer to \cite{Adamek:2017uiq}.}}
Other cosmological parameters are summarized in Table \ref{Table1}.
From each snapshot, we construct 64 sub-boxes of side $ 512 h^{-1} {\rm Mpc} $ and find the cosmic environments. Figure \ref{fig:Env-Autumn} shows the results for a $ 5 h^{-1} {\rm Mpc}$  thick slice of a sub-box from $ \Lambda$CDM simulation box. 

\subsection{Cosmic Web Classification}\label{sec:CW-Classif}
\begin{figure}
\centering
\includegraphics[width=0.48\textwidth]{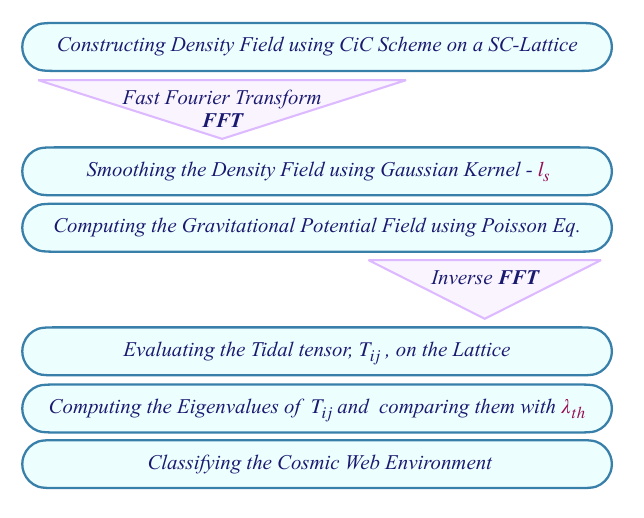}
\caption{The workflow of the cosmic environments classification. The process starts by constructing the density field $ \rho (x) $ on a simple cubic lattice (SC-lattice) using the coordinates and masses of the halos in the catalogue and ends by specifying the environment in which a halo resides. The lattice parameter---$l_p$, the smoothing scale---$ l_s$, and the threshold value---$\lambda_{th}$ for the eigenvalues of the $ T_{ij}$ tensor are three parameters introduced in this process. } 
\label{fig:WorkFlow}
\end{figure}

In this paper, we use the T-web classification method introduced and developed by \cite{Hahn:2007ui,  Forero-Romero:2008svv}. In Figure \ref{fig:WorkFlow}, we summarize the workflow of the classification process. 
This process starts by constructing the density field $\rho(\textbf{x})$ on a simple cubic lattice, with a lattice parameter $ l_p $, from the coordinates and mass of halos using the cloud-in-cell (CiC) scheme. 
Then, we compute the density contrast field at each lattice point by $\delta(\textbf{x})=\rho(\textbf{x})/\bar\rho-1$.  After transforming the density contrast field to the Fourier space, we smooth the field by a Gaussian window function with a smoothing scale of $l_s$. Later, we compute the normalized gravitational potential using the normalized Poisson equation, $ \nabla^2 \Phi = 4\pi G\bar\rho \  \delta $.  After the inverse Fourier transforms, we calculate the T-web tensor,  equation \ref{T-tensor}, at each grid and find the eigenvalues of it at each lattice point.
Then, we determine the environment of each lattice point by comparing the eigenvalues with the $\lambda_{th}$ with the criteria mentioned in Section \ref{sec:CWE}. 
\\
Three parameters---$l_p$, $l_s$, and $\lambda_{th}$---introduced in this process. The typical interspacing between halos and the computation cost limits the choice for $l_p$. Based on these limits, we construct a lattice with $300^3$ points in each of the 64 sub-boxes for each simulation snapshot. This choice sets the $ l_p$ around $ 1.6 \  h^{-1} {\rm Mpc} $. This value is in the same order as the peak of the first nearest neighbour probability distribution function. 

%%****************FIG 2**********************%%
%%****************FIG 2**********************%%

%%****************FIG 3**********************%%
\begin{figure}
\centering
\includegraphics[width=0.48\textwidth]{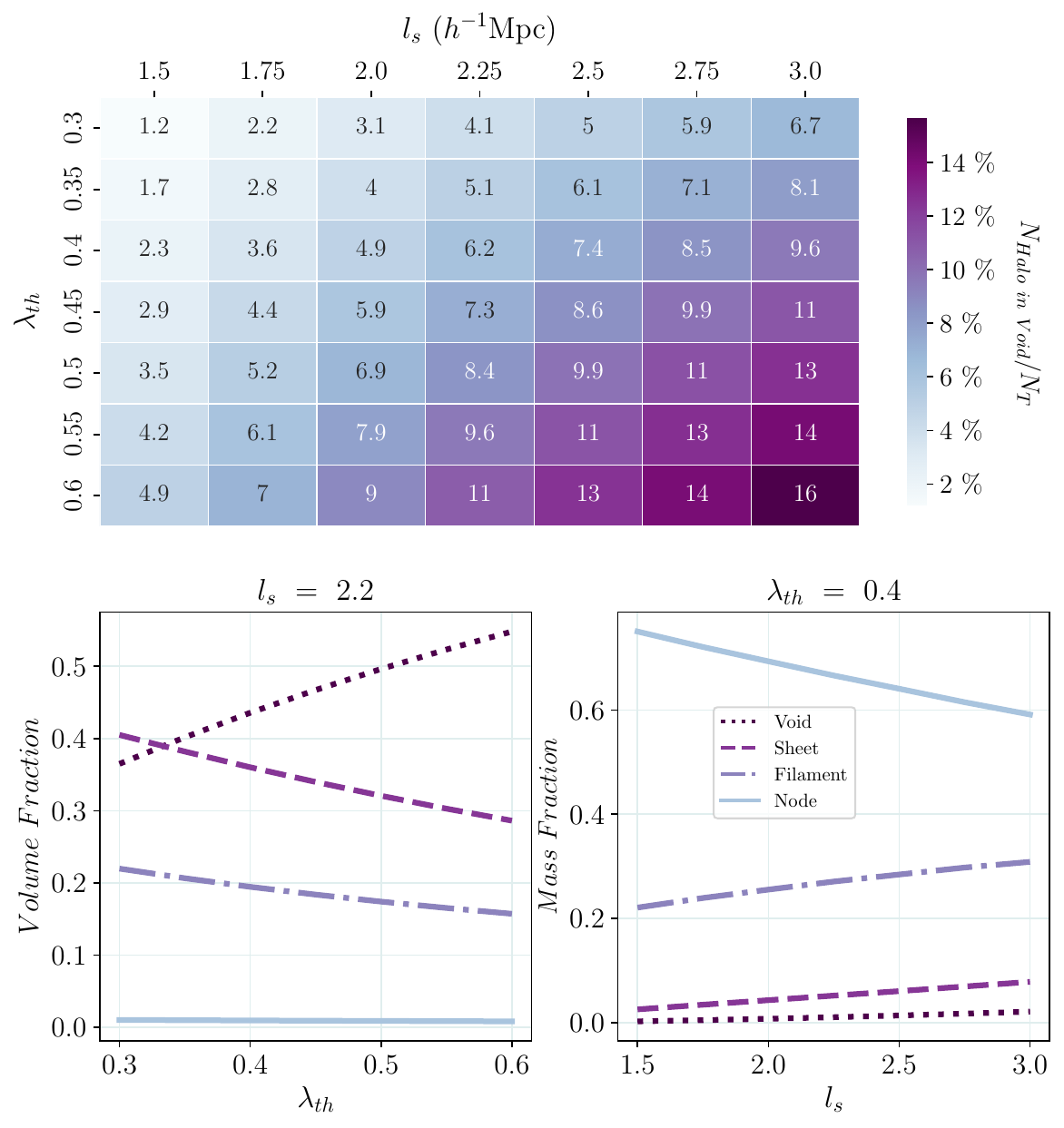}
\caption{Studying the effect of change of the smoothing scale--$l_s $ and $\lambda_{th}$ on the environments. The top panel shows the heat map of the percentage of halo residue in the voids when the two parameters vary. The bottom left plot indicates the volume fraction of various environments for fixed $ l_s =2.2 \  h^{-1} {\rm Mpc} $ and different $ \lambda_{th}$. The bottom right plot shows the mass fraction of various environments for fixed $ \lambda_{th} = 0.4$ and different $ l_s$. In the bottom plots, the solid and dash-dotted lines stand for nodes and filaments while the dotted and dash lines present void and sheet, respectively.} 
\label{fig:Void-halo}
\end{figure}
%%****************FIG 3**********************%%

The choice for ($l_s$, $\lambda_{th}$) is a bit challenging. To find the suitable values for these two parameters, we run the environment classification process several times by changing them slightly at each run. Figure \ref{fig:Void-halo} summarizes the results. The heat map on the top of Figure \ref{fig:Void-halo} indicates the percentage of the halos belonging to voids concerning all halos. Our choice of parameters should allow for a percentage that provides statistical significance for our analysis. We pick (0.4,  2.2) for ($l_s$, $\lambda_{th}$), with these at least five percent of halos belonging to the voids. To check the physical rationality behind this choice of parameters, we adjust one parameter while allowing the other parameter to vary within a specific interval.
With $ l_s = 2 $, the choice for $ \lambda_{th}$ must be in the region where the hierarchy of the volume fraction of various environments stays the same, the bottom left plot of Figure \ref{fig:Void-halo}.
For $ \lambda_{th}=0.4$, the value for the $ l_s$ should be above the points where the mass fraction of various environments experience a considerable change, the bottom right of Figure \ref{fig:Void-halo}.
\\
In Appendix \ref{App_1}, we study the impact of varying the $ \lambda_{th}$ and $ l_{s}$ on the statistics used in this work. 

\section{Results :Letter functions and environment}\label{Sec4}

%%****************FIG 4**********************%%
\begin{figure}
\centering
\includegraphics[width=0.48\textwidth]{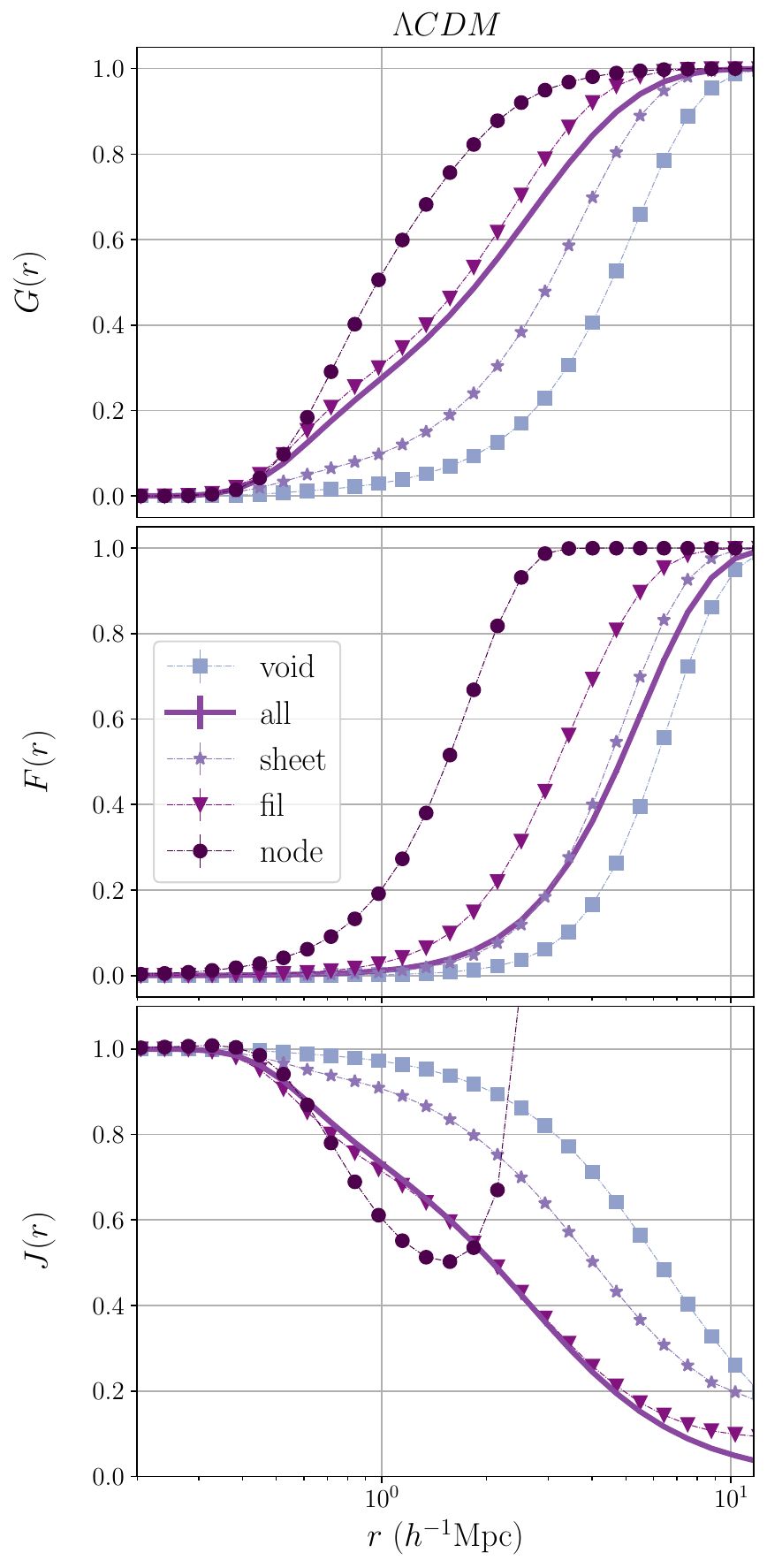}
\caption{Letter functions computed in various cosmic environments for $\Lambda$CDM model, from top to bottom $G$, $F$, and $J$ functions. In all panels, the solid line shows the behaviour of all the data points, and the squares, stars, triangles, and circles represent the points belonging to the voids, sheets, filaments, and nodes, respectively. The error bars are smaller than the markers of the data points.}
\label{fig:G-Env-LCDM}
\end{figure}
%%****************FIG 4**********************%%

In the following section, we study different cosmic environments using the letter functions---$G$ or the nearest neighbour CDF, the $ F$ or the spherical contact CDF, and $J$-function as defined by equation \ref{Eq-Jfun}---in the $\Lambda$CDM model and cosmologies with massive neutrinos. 
As explained in Section 3, we divide each simulation snapshot into 64 sub-boxes of side 512 $h^{-1} \rm{Mpc}$ and run the environment classification algorithm on each of these 64 sub-boxes, and then compute our favoured quantities (e.g., $G$ in voids) in each of these sub-boxes. In the Figures \ref{fig:G-Env-LCDM} and  \ref{fig:LFs-Env},  the data points and the error bars are the mean and standard deviation of 64 values from each sub-box.
\\

\subsection{$\Lambda$CDM model}
Figure \ref{fig:G-Env-LCDM} shows the letter functions computed in various environments within the $ \Lambda$CDM model. The solid line represents the behaviour of all the points regardless of their habitats. Squares and stars stand for the points residue in the voids and on the sheets; Triangles and circles represent the points in the filaments and the nodes. 
The top panel of Figure \ref{fig:G-Env-LCDM} shows the $G$ function---the CDF of the first nearest neighbour for the halos. It is possible to make the following observations from this diagram:
\begin{itemize}
\item The $ G_{all}$ function follows mostly the $ G$ for filamentary halos (i.e., the triangles). This behaviour is rooted in the fact that the filamentary halos outnumbered all the halos in the other three environments. \cite{GaneshaiahVeena:2020bsx}  among many others, pointed out that most halos in the Universe live in filaments. 
\item The $ G $ functions of less dense environments (i.e., stars for sheets and squares for voids) are below the $  G_{all}$ for all distances, which translates to less probability of finding the immediate neighbour in these environments.
\item The $ G$ functions of extreme environments, node (i.e., circles), and void  (i.e., squares) envelope the $G$ of other environments. In addition, at the scale that almost all node halos find their immediate neighbours, most of the void halos are still at the centre of an empty sphere without any halos on the boundary.
\end{itemize}
The middle panel of Figure \ref{fig:G-Env-LCDM}  indicates the $ F$ function---the CDF of the distance to the first nearest halos from randomly generated points. To calculate this, first, we prepare a randomly generated set of points in the sub-box, then determine the environments for each random point, and finally find the distance to the nearest neighbouring halo (regardless of the environment that the halo belongs to). Observations from the diagram:
\begin{itemize}
\item The $ F_{all}$ tends toward the $ F$ functions of sheets and voids since these environments are voluminous and contain the most randomly generated points. 
\item The $ F$ functions of the extreme environments envelope the other $F$ function.
\end{itemize}
In the bottom panel of Figure  \ref{fig:G-Env-LCDM}, we plot the $ J$-function. As mentioned in Section \ref{1-pointstat}, the $ J$-function quantifies the extent of the clustering of a point pattern when compared with unity. A significant deviation from the unity indicates a high degree of clustering in the point pattern. In this diagram, as anticipated, voids are less clustered environments, while nodes are highly clustered.  
It is worth noting that the turnover in the $ J$-function is because the denominator of the equation \ref{Eq-Jfun},  $1 - F(r)$,  becomes zero.
\\
Figure \ref{fig:G-Env-LCDM} shows that the letter functions can differentiate between different environments.

\subsection{$\nu \Lambda$CDM Versus $\Lambda$CDM}

\begin{figure*}
\centering
\includegraphics[width=1\textwidth]{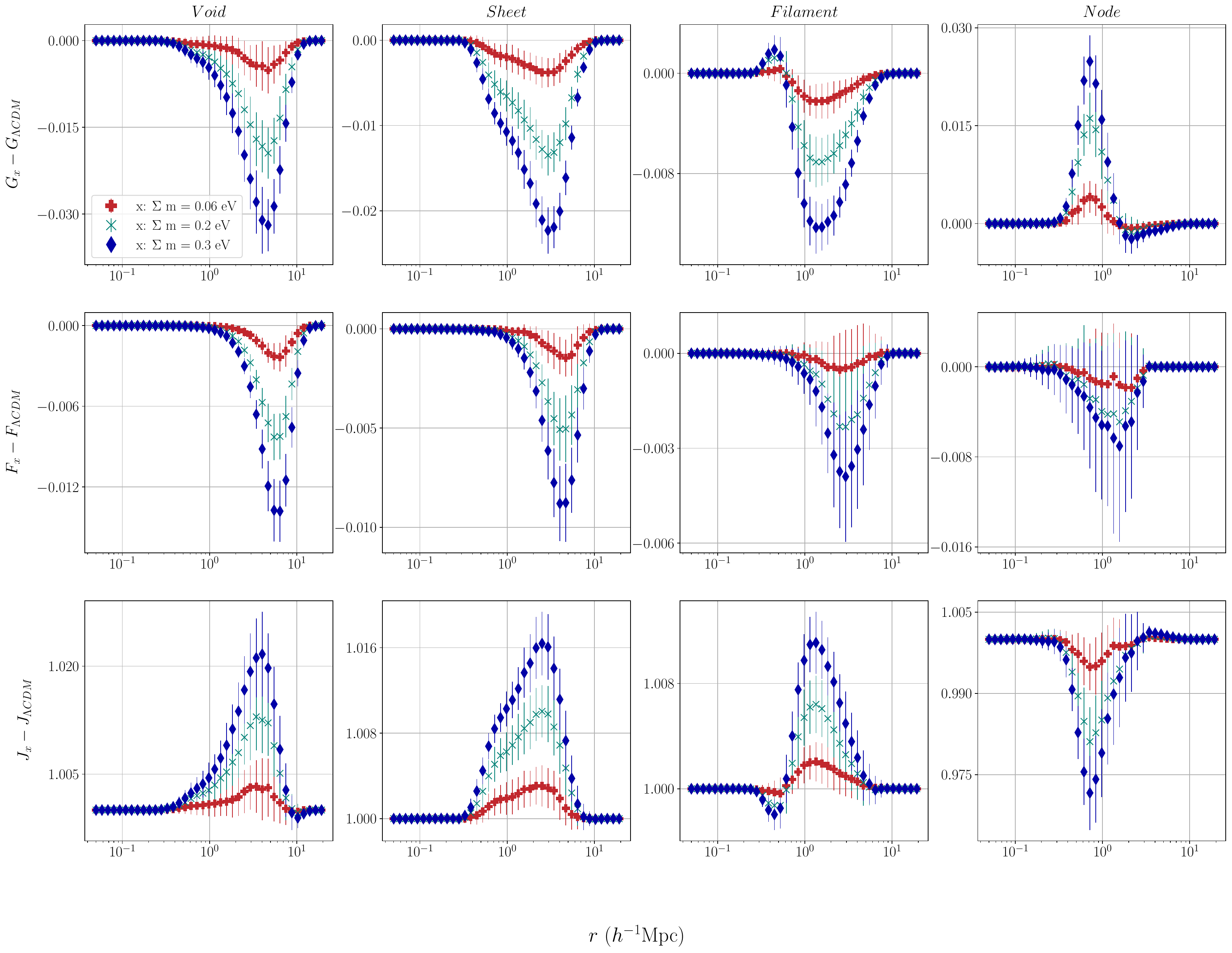}
\caption{The difference between letter functions in $\Lambda$CDM model and cosmologies with massive neutrinos. The plus, cross, and diamond represent cosmologies with a total neutrino mass of $\Sigma \ m_{\nu} = 0.06, 0.2, 0.3 $ eV, respectively.} 
\label{fig:LFs-Env}
\end{figure*}

In Figure \ref{fig:LFs-Env}, we plot the difference in letter functions between the $ \Lambda$CDM model and cosmologies with massive neutrinos. In this plot, plus, cross, and diamond indicate the total neutrino mass of $0.06, 0.2, 0.3$ eV.
Looking at this diagram, one can note the following:
\begin{itemize}
\item The $G$ function can distinguish between different total masses in every environment. The deviation from the $ \Lambda$CDM model increases as the total neutrino mass increases. The degree of deviation from the standard model decreases as the cosmic environment becomes more dense.
\item {{The $ F$ function can distinguish between different total neutrino masses in the voluminous environment. In dense and compact environments, i.e., filaments and nodes, there is a noticeable trend depending on the total neutrino mass. However, because of the large error bars, the $ F$ function can not differentiate between the total neutrino masses. Starting with larger volumes (i.e., bigger sub-boxes) can improve the situation by decreasing the errors due to the sample variance. Then, one can anticipate that the $ F$ function can differentiate between various total neutrino masses, even in filaments and nodes.}}

\item The $ J$-function indicates that as the total neutrino mass increases, the extent of clustering decreases as opposed to the $ \Lambda$CDM model.
\end{itemize}

By investigating the samples using letter functions, one can cross-check the presence of the massive neutrino mass. The presence of this deviation in all three is the sanity of the method. If the effect were absent in one of the functions,  this would be an alarm for the model. 
In the recent work by \cite{Khoshtinat:2023zck}, we showed that the letter functions can distinguish between different cosmologies. 
We demonstrate in this paper that the letter functions computed in distinct cosmic environments can serve as complementary tools to differentiate between various cosmological models.

%%****************FIG 5**********************%%
%%****************FIG 5**********************%%

\section{Conclusions and Future remarks}\label{Sec5}
The distribution of galaxies contains information that can constrain and distinguish cosmological models. Galaxies reside inside dark matter halos, which make the cosmic skeleton known as the cosmic web. The cosmic web consists of different elements: nodes, filaments, sheets, and voids. In this work, we compute the one-point statistics---$G$, $F$, and $J$ functions---in different cosmic environments for a suit of the gevolution simulation. We classify the cosmic environments via the T-web method, described in Section \ref{sec:CWE}. 

In the first step, we study the standard model and show that the letter functions can distinguish the different environments. $G$ and $ J$ functions for all halos mimic the results obtained from the halos in filaments. However, in the case of the $F$ function, the whole sample resembles the sheet. One step further, we study these functions in cosmologies with massive neutrinos. We propose that the letter functions computed for different cosmic environments fit as complementary tools for differentiating the extension of the standard model. In the case of cosmologies with massive neutrinos, the voluminous underdense regions--voids and sheets--are more prominent for doing this task. 

In the T-web algorithm, the smoothing scale $l_s$ of the density field and the threshold criteria for eigenvalues of T-web $\lambda_{th}$ are the two main parameters of the environment-finding algorithm. To find the plausible values for  $l_s$ and $\lambda_{th}$, we repeat the algorithm for several values of $l_s$  and $\lambda_{th}$ and based on the heat map for the number of halos in voids and the behaviour of the volume and mass fractions choose  $2.2 \ h^{-1} \rm{Mpc}$ and $0.4$ for $l_s$ and  $\lambda_{th}$, respectively.\\

{{For future studies, we have two directions in our mind: \begin{itemize}
\item First,  we should examine the one-point statistics in the other extensions of the $ \Lambda$CDM model, such as dark energy models, modified dark matter models, and also in a mixed scenario. It is critical because it allows us to study the possible degeneracies in the one-point statistics between different parameters. 
\item Second, we should build a bridge between our results from the simulations and the results from the upcoming surveys. For this task, in the first step, one should run a large set of simulations for each free parameter, e.g., the total neutrino mass,  to construct an emulator based on the one-point statistics we used (a quality example would be \cite{Parimbelli:2021mtp}, where they studied a mixed dark matter scenario). The next stage is constructing the mock galaxy catalogues, starting with halo catalogues using the physics of the halo occupation distribution. At this stage, one can run the environment identification algorithm and equip the emulator with further information from the different cosmic environments. Then, the output is a reasonable theoretical prediction of what we expect to get from galaxy survey observations. Concerning the data from the galaxy surveys, the environment identification algorithm can run on the volume and flux-limited catalogues. In other words, one can consider the galaxies as a point-wise particle in the observational box and use the proper mass-to-light ratio to construct the density field, which is the input of the environmental classification algorithm ( an example of studying specific environment is \cite{Tavasoli:2021reo}, the authors used a graph-based method to study the void galaxies from the cosmological surveys as well as the mock catalogues).
\end{itemize}}}

\section*{Acknowledgements}
We are very grateful to Farbod Hassani,  Julian Adamek, and Martin Kunz for giving us access to the halo catalogues of high-resolution neutrino simulations presented in \citep{Adamek:2017uiq}. 
SB is partially supported by the Abdus Salam International Centre for Theoretical Physics (ICTP) under the regular associateship scheme.
Moreover, MK and SB are partially supported by the Sharif University of Technology Office of Vice President for Research under Grant No. G4010204.

%%%%%%%%%%%%%%%%%%%%%%%%%%%%%%%%%%%%%%%%%%%%%%%%%%
\section*{Data Availability}
The halo catalogues we used are based on the gevolution simulation \citep{Adamek:2017uiq}. 

%%%%%%%%%%%%%%%%%%%%%%%%%%%%%%%%%%%%%%%%%%%%%%%%%%



\bibliographystyle{mnras}
\bibliography{Bib.bib} % if your bibtex file is called example.bib

\begin{thebibliography}{}
\makeatletter
\relax
\def\mn@urlcharsother{\let\do\@makeother \do\$\do\&\do\#\do\^\do\_\do\%\do\~}
\def\mn@doi{\begingroup\mn@urlcharsother \@ifnextchar [ {\mn@doi@}
  {\mn@doi@[]}}
\def\mn@doi@[#1]#2{\def\@tempa{#1}\ifx\@tempa\@empty \href
  {http://dx.doi.org/#2} {doi:#2}\else \href {http://dx.doi.org/#2} {#1}\fi
  \endgroup}
\def\mn@eprint#1#2{\mn@eprint@#1:#2::\@nil}
\def\mn@eprint@arXiv#1{\href {http://arxiv.org/abs/#1} {{\tt arXiv:#1}}}
\def\mn@eprint@dblp#1{\href {http://dblp.uni-trier.de/rec/bibtex/#1.xml}
  {dblp:#1}}
\def\mn@eprint@#1:#2:#3:#4\@nil{\def\@tempa {#1}\def\@tempb {#2}\def\@tempc
  {#3}\ifx \@tempc \@empty \let \@tempc \@tempb \let \@tempb \@tempa \fi \ifx
  \@tempb \@empty \def\@tempb {arXiv}\fi \@ifundefined
  {mn@eprint@\@tempb}{\@tempb:\@tempc}{\expandafter \expandafter \csname
  mn@eprint@\@tempb\endcsname \expandafter{\@tempc}}}

\bibitem[\protect\citeauthoryear{Abazajian et~al.}{Abazajian
  et~al.}{2009}]{SDSS:2008tqn}
Abazajian K.~N.,  et~al., 2009, \mn@doi [Astrophys. J. Suppl.]
  {10.1088/0067-0049/182/2/543}, 182, 543

\bibitem[\protect\citeauthoryear{Adame et~al.}{Adame
  et~al.}{2023}]{DESI:2023ytc}
Adame G.,  et~al., 2023, arXiv:2306.06308

\bibitem[\protect\citeauthoryear{Adamek, Daverio, Durrer  \& Kunz}{Adamek
  et~al.}{2016a}]{Adamek:2016zes}
Adamek J.,  Daverio D.,  Durrer R.,   Kunz M.,  2016a, \mn@doi [JCAP]
  {10.1088/1475-7516/2016/07/053}, 07, 053

\bibitem[\protect\citeauthoryear{Adamek, Daverio, Durrer  \& Kunz}{Adamek
  et~al.}{2016b}]{Adamek:2015eda}
Adamek J.,  Daverio D.,  Durrer R.,   Kunz M.,  2016b, \mn@doi [Nature Physics]
  {10.1038/nphys3673}, 12, 346

\bibitem[\protect\citeauthoryear{Adamek, Durrer  \& Kunz}{Adamek
  et~al.}{2017}]{Adamek:2017uiq}
Adamek J.,  Durrer R.,   Kunz M.,  2017, \mn@doi [JCAP]
  {10.1088/1475-7516/2017/11/004}, 11, 004

\bibitem[\protect\citeauthoryear{Agarwal \& Feldman}{Agarwal \&
  Feldman}{2011}]{Agarwal:2010mt}
Agarwal S.,  Feldman H.~A.,  2011, \mn@doi [Mon. Not. Roy. Astron. Soc.]
  {10.1111/j.1365-2966.2010.17546.x}, 410, 1647

\bibitem[\protect\citeauthoryear{Aghanim et~al.}{Aghanim
  et~al.}{2020}]{Planck:2018vyg}
Aghanim N.,  et~al., 2020, \mn@doi [Astron. Astrophys.]
  {10.1051/0004-6361/201833910}, 641, A6

\bibitem[\protect\citeauthoryear{Aker et~al.}{Aker
  et~al.}{2022}]{KATRIN:2021uub}
Aker M.,  et~al., 2022, \mn@doi [Nature Phys.] {10.1038/s41567-021-01463-1},
  18, 160

\bibitem[\protect\citeauthoryear{Almeida et~al.}{Almeida
  et~al.}{2023}]{SDSS:2023tbz}
Almeida A.,  et~al., 2023, \mn@doi [Astrophys. J. Suppl.]
  {10.3847/1538-4365/acda98}, 267, 44

\bibitem[\protect\citeauthoryear{Amendola}{Amendola}{2000}]{Amendola:1999er}
Amendola L.,  2000, \mn@doi [Phys. Rev. D] {10.1103/PhysRevD.62.043511}, 62,
  043511

\bibitem[\protect\citeauthoryear{Amendola \& Tsujikawa}{Amendola \&
  Tsujikawa}{2015}]{Amendola:2015ksp}
Amendola L.,  Tsujikawa S.,  2015, {\it{{Dark Energy}: {Theory and
  Observations}}}.
Cambridge University Press

\bibitem[\protect\citeauthoryear{Arnold, Shandarin  \& Zeldovich}{Arnold
  et~al.}{1982}]{arnold1982large}
Arnold V.,  Shandarin S.,   Zeldovich Y.~B.,  1982, Geophysical \&
  Astrophysical Fluid Dynamics, 20, 111

\bibitem[\protect\citeauthoryear{Ay\c{c}oberry, Barthelemy  \&
  Codis}{Ay\c{c}oberry et~al.}{2023}]{Aycoberry:2023uyk}
Ay\c{c}oberry E.,  Barthelemy A.,   Codis S.,  2023, arXiv:2310.03548

\bibitem[\protect\citeauthoryear{Banerjee \& Abel}{Banerjee \&
  Abel}{2020}]{Banerjee:2020umh}
Banerjee A.,  Abel T.,  2020, \mn@doi [Mon. Not. Roy. Astron. Soc.]
  {10.1093/mnras/staa3604}, 500, 5479

\bibitem[\protect\citeauthoryear{Banerjee, Kokron  \& Abel}{Banerjee
  et~al.}{2022}]{Banerjee:2021cmi}
Banerjee A.,  Kokron N.,   Abel T.,  2022, \mn@doi [Mon. Not. Roy. Astron.
  Soc.] {10.1093/mnras/stac193}, 511, 2765

\bibitem[\protect\citeauthoryear{Bardeen, Bond, Kaiser  \& Szalay}{Bardeen
  et~al.}{1986}]{Bardeen:1985tr}
Bardeen J.~M.,  Bond J.~R.,  Kaiser N.,   Szalay A.~S.,  1986, \mn@doi
  [Astrophys. J.] {10.1086/164143}, 304, 15

\bibitem[\protect\citeauthoryear{Bartolo, Komatsu, Matarrese  \&
  Riotto}{Bartolo et~al.}{2004}]{Bartolo:2004if}
Bartolo N.,  Komatsu E.,  Matarrese S.,   Riotto A.,  2004, \mn@doi [Phys.
  Rept.] {10.1016/j.physrep.2004.08.022}, 402, 103

\bibitem[\protect\citeauthoryear{Behroozi, Wechsler  \& Wu}{Behroozi
  et~al.}{2013}]{Behroozi:2011ju}
Behroozi P.~S.,  Wechsler R.~H.,   Wu H.-Y.,  2013, \mn@doi [Astrophys. J.]
  {10.1088/0004-637X/762/2/109}, 762, 109

\bibitem[\protect\citeauthoryear{Bernardeau, Colombi, Gaztanaga  \&
  Scoccimarro}{Bernardeau et~al.}{2002}]{Bernardeau:2001qr}
Bernardeau F.,  Colombi S.,  Gaztanaga E.,   Scoccimarro R.,  2002, \mn@doi
  [Phys. Rept.] {10.1016/S0370-1573(02)00135-7}, 367, 1

\bibitem[\protect\citeauthoryear{Bond, Cole, Efstathiou  \& Kaiser}{Bond
  et~al.}{1991}]{Bond:1990iw}
Bond J.~R.,  Cole S.,  Efstathiou G.,   Kaiser N.,  1991, \mn@doi [Astrophys.
  J.] {10.1086/170520}, 379, 440

\bibitem[\protect\citeauthoryear{Bonnaire, Aghanim, Kuruvilla  \&
  Decelle}{Bonnaire et~al.}{2022}]{Bonnaire:2021sie}
Bonnaire T.,  Aghanim N.,  Kuruvilla J.,   Decelle A.,  2022, \mn@doi [Astron.
  Astrophys.] {10.1051/0004-6361/202142852}, 661, A146

\bibitem[\protect\citeauthoryear{Bonnaire, Kuruvilla, Aghanim  \&
  Decelle}{Bonnaire et~al.}{2023}]{Bonnaire:2022ocm}
Bonnaire T.,  Kuruvilla J.,  Aghanim N.,   Decelle A.,  2023, \mn@doi [Astron.
  Astrophys.] {10.1051/0004-6361/202245626}, 674, A150

\bibitem[\protect\citeauthoryear{Colberg, Sheth, Diaferio, Gao  \&
  Yoshida}{Colberg et~al.}{2005}]{Colberg:2004nd}
Colberg J.~M.,  Sheth R.~K.,  Diaferio A.,  Gao L.,   Yoshida N.,  2005,
  \mn@doi [Mon. Not. Roy. Astron. Soc.] {10.1111/j.1365-2966.2005.09064.x},
  360, 216

\bibitem[\protect\citeauthoryear{Cooray \& Sheth}{Cooray \&
  Sheth}{2002}]{Cooray:2002dia}
Cooray A.,  Sheth R.~K.,  2002, \mn@doi [Phys. Rept.]
  {10.1016/S0370-1573(02)00276-4}, 372, 1

\bibitem[\protect\citeauthoryear{Di~Valentino et~al.,}{Di~Valentino
  et~al.}{2021}]{di2021realm}
Di~Valentino E.,  et~al., 2021, Classical and Quantum Gravity, 38, 153001

\bibitem[\protect\citeauthoryear{Doroshkevich \& Shandarin}{Doroshkevich \&
  Shandarin}{1978}]{doroshkevich1978statistical}
Doroshkevich A.,  Shandarin S.,  1978, Soviet Astronomy, vol. 22, Nov.-Dec.
  1978, p. 653-660. Translation. Astronomicheskii Zhurnal, vol. 55, Nov.-Dec.
  1978, p. 1144-1156., 22, 653

\bibitem[\protect\citeauthoryear{Dournac et~al.}{Dournac
  et~al.}{2024}]{Euclid:2024vss}
Dournac F.,  et~al., 2024, arXiv:2404.12157

\bibitem[\protect\citeauthoryear{Fard, Taamoli  \& Baghram}{Fard
  et~al.}{2019}]{Fard:2018dwx}
Fard M.~A.,  Taamoli S.,   Baghram S.,  2019, \mn@doi [Mon. Not. Roy. Astron.
  Soc.] {10.1093/mnras/stz2210}, 489, 900

\bibitem[\protect\citeauthoryear{Fard, Baghkhani, Ghodsi, Taamoli, Hassani  \&
  Baghram}{Fard et~al.}{2022}]{Fard:2021qaa}
Fard M.~A.,  Baghkhani Z.,  Ghodsi L.,  Taamoli S.,  Hassani F.,   Baghram S.,
  2022, \mn@doi [Mon. Not. Roy. Astron. Soc.] {10.1093/mnras/stac256}, 512,
  5165

\bibitem[\protect\citeauthoryear{Forero-Romero, Hoffman, Gottloeber, Klypin  \&
  Yepes}{Forero-Romero et~al.}{2009}]{Forero-Romero:2008svv}
Forero-Romero J.~E.,  Hoffman Y.,  Gottloeber S.,  Klypin A.,   Yepes G.,
  2009, \mn@doi [Mon. Not. Roy. Astron. Soc.]
  {10.1111/j.1365-2966.2009.14885.x}, 396, 1815

\bibitem[\protect\citeauthoryear{Ganeshaiah~Veena, Cautun, van~de Weygaert,
  Tempel  \& Frenk}{Ganeshaiah~Veena et~al.}{2021}]{GaneshaiahVeena:2020bsx}
Ganeshaiah~Veena P.,  Cautun M.,  van~de Weygaert R.,  Tempel E.,   Frenk
  C.~S.,  2021, \mn@doi [Mon. Not. Roy. Astron. Soc.] {10.1093/mnras/stab411},
  503, 2280

\bibitem[\protect\citeauthoryear{Gruen et~al.}{Gruen
  et~al.}{2018}]{DES:2017eav}
Gruen D.,  et~al., 2018, \mn@doi [Phys. Rev. D] {10.1103/PhysRevD.98.023507},
  98, 023507

\bibitem[\protect\citeauthoryear{Hahn, Carollo, Porciani  \& Dekel}{Hahn
  et~al.}{2007}]{Hahn:2007ui}
Hahn O.,  Carollo C.~M.,  Porciani C.,   Dekel A.,  2007, \mn@doi [Mon. Not.
  Roy. Astron. Soc.] {10.1111/j.1365-2966.2007.12249.x}, 381, 41

\bibitem[\protect\citeauthoryear{Hand}{Hand}{2008}]{hand2008statistical}
Hand D.~J.,  2008, {\it{Statistical analysis and modelling of spatial point
  patterns}} by Janine Illian, Antti Penttinen, Helga Stoyan, Dietrich Stoyan

\bibitem[\protect\citeauthoryear{Hannestad, Upadhye  \& Wong}{Hannestad
  et~al.}{2020}]{Hannestad:2020rzl}
Hannestad S.,  Upadhye A.,   Wong Y. Y.~Y.,  2020, \mn@doi [JCAP]
  {10.1088/1475-7516/2020/11/062}, 11, 062

\bibitem[\protect\citeauthoryear{Huchra, Davis, Latham  \& Tonry}{Huchra
  et~al.}{1983}]{Huchra:1983wy}
Huchra J.,  Davis M.,  Latham D.,   Tonry J.,  1983, \mn@doi [Astrophys. J.
  Suppl.] {10.1086/190860}, 52, L89

\bibitem[\protect\citeauthoryear{Joudaki et~al.}{Joudaki
  et~al.}{2018}]{Joudaki:2017zdt}
Joudaki S.,  et~al., 2018, \mn@doi [Mon. Not. Roy. Astron. Soc.]
  {10.1093/mnras/stx2820}, 474, 4894

\bibitem[\protect\citeauthoryear{Kaiser}{Kaiser}{1987}]{Kaiser:1987qv}
Kaiser N.,  1987, \mn@doi [Mon. Not. Roy. Astron. Soc.]
  {10.1093/mnras/227.1.1}, 227, 1

\bibitem[\protect\citeauthoryear{Khoshtinat, Ansarifard, Hassani  \&
  Baghram}{Khoshtinat et~al.}{2024}]{Khoshtinat:2023zck}
Khoshtinat M.,  Ansarifard M.,  Hassani F.,   Baghram S.,  2024, \mn@doi [Mon.
  Not. Roy. Astron. Soc.] {10.1093/mnras/stae1195}, 531, 575

\bibitem[\protect\citeauthoryear{Kousha, Ansarifard  \& Abolhasani}{Kousha
  et~al.}{2024}]{Kousha:2023kog}
Kousha H.~M.,  Ansarifard M.,   Abolhasani A.,  2024, \mn@doi [Mon. Not. Roy.
  Astron. Soc.] {10.1093/mnras/stae1631}, 532, 2356

\bibitem[\protect\citeauthoryear{Lesgourgues \& Pastor}{Lesgourgues \&
  Pastor}{2006}]{Lesgourgues:2006nd}
Lesgourgues J.,  Pastor S.,  2006, \mn@doi [Phys. Rept.]
  {10.1016/j.physrep.2006.04.001}, 429, 307

\bibitem[\protect\citeauthoryear{Libeskind et~al.}{Libeskind
  et~al.}{2018}]{Libeskind:2017tun}
Libeskind N.~I.,  et~al., 2018, \mn@doi [Mon. Not. Roy. Astron. Soc.]
  {10.1093/mnras/stx1976}, 473, 1195

\bibitem[\protect\citeauthoryear{Massara, Villaescusa-Navarro, Ho, Dalal  \&
  Spergel}{Massara et~al.}{2021}]{Massara:2020pli}
Massara E.,  Villaescusa-Navarro F.,  Ho S.,  Dalal N.,   Spergel D.~N.,  2021,
  \mn@doi [Phys. Rev. Lett.] {10.1103/PhysRevLett.126.011301}, 126, 011301

\bibitem[\protect\citeauthoryear{Navarro, Frenk  \& White}{Navarro
  et~al.}{1996}]{Navarro:1995iw}
Navarro J.~F.,  Frenk C.~S.,   White S. D.~M.,  1996, \mn@doi [Astrophys. J.]
  {10.1086/177173}, 462, 563

\bibitem[\protect\citeauthoryear{Parimbelli, Scelfo, Giri, Schneider,
  Archidiacono, Camera  \& Viel}{Parimbelli et~al.}{2021}]{Parimbelli:2021mtp}
Parimbelli G.,  Scelfo G.,  Giri S.~K.,  Schneider A.,  Archidiacono M.,
  Camera S.,   Viel M.,  2021, \mn@doi [JCAP] {10.1088/1475-7516/2021/12/044},
  12, 044

\bibitem[\protect\citeauthoryear{Parkavousi, Kameli  \& Baghram}{Parkavousi
  et~al.}{2023}]{Parkavousi:2022llz}
Parkavousi L.,  Kameli H.,   Baghram S.,  2023, \mn@doi [Mon. Not. Roy. Astron.
  Soc.] {10.1093/mnras/stad2829}, 526, 1495

\bibitem[\protect\citeauthoryear{Peebles \& Ratra}{Peebles \&
  Ratra}{2003}]{Peebles:2002gy}
Peebles P. J.~E.,  Ratra B.,  2003, \mn@doi [Rev. Mod. Phys.]
  {10.1103/RevModPhys.75.559}, 75, 559

\bibitem[\protect\citeauthoryear{Perivolaropoulos \& Skara}{Perivolaropoulos \&
  Skara}{2022}]{Perivolaropoulos:2021jda}
Perivolaropoulos L.,  Skara F.,  2022, \mn@doi [New Astron. Rev.]
  {10.1016/j.newar.2022.101659}, 95, 101659

\bibitem[\protect\citeauthoryear{Ravi~K}{Ravi~K}{1996}]{ravi1996distribution}
Ravi~K S.,  1996, Mon. Not. Roy. Astron. Soc., 281, 1124

\bibitem[\protect\citeauthoryear{Sheth \& Tormen}{Sheth \&
  Tormen}{2004}]{Sheth:2004vb}
Sheth R.~K.,  Tormen G.,  2004, \mn@doi [Mon. Not. Roy. Astron. Soc.]
  {10.1111/j.1365-2966.2004.07733.x}, 350, 1385

\bibitem[\protect\citeauthoryear{Sheth \& van~de Weygaert}{Sheth \& van~de
  Weygaert}{2004}]{Sheth:2003py}
Sheth R.~K.,  van~de Weygaert R.,  2004, \mn@doi [Mon. Not. Roy. Astron. Soc.]
  {10.1111/j.1365-2966.2004.07661.x}, 350, 517

\bibitem[\protect\citeauthoryear{Springel}{Springel}{2005}]{Springel:2005mi}
Springel V.,  2005, \mn@doi [Mon. Not. Roy. Astron. Soc.]
  {10.1111/j.1365-2966.2005.09655.x}, 364, 1105

\bibitem[\protect\citeauthoryear{Stoughton et~al.}{Stoughton
  et~al.}{2002}]{SDSS:2002oin}
Stoughton C.,  et~al., 2002, \mn@doi [Astron. J.] {10.1086/324741}, 123, 485

\bibitem[\protect\citeauthoryear{Stoyan}{Stoyan}{1984}]{stoyan1984correlations}
Stoyan D.,  1984, Mathematische Nachrichten, 116, 197

\bibitem[\protect\citeauthoryear{Tabatabaei, Banihashemi, Baghram  \&
  Mashhoon}{Tabatabaei et~al.}{2023}]{Tabatabaei:2023qxw}
Tabatabaei J.,  Banihashemi A.,  Baghram S.,   Mashhoon B.,  2023, \mn@doi
  [Int. J. Mod. Phys. D] {10.1142/S0218271823420099}, 32, 2342009

\bibitem[\protect\citeauthoryear{Tavasoli}{Tavasoli}{2021}]{Tavasoli:2021reo}
Tavasoli S.,  2021, \mn@doi [Astrophys. J. Lett.] {10.3847/2041-8213/ac1357},
  916, L24

\bibitem[\protect\citeauthoryear{Uhlemann, Friedrich, Villaescusa-Navarro,
  Banerjee  \& Codis}{Uhlemann et~al.}{2020}]{Uhlemann:2019gni}
Uhlemann C.,  Friedrich O.,  Villaescusa-Navarro F.,  Banerjee A.,   Codis S.,
  2020, \mn@doi [Mon. Not. Roy. Astron. Soc.] {10.1093/mnras/staa1155}, 495,
  4006

\bibitem[\protect\citeauthoryear{Uhlemann, Friedrich, Boyle, Gough, Barthelemy,
  Bernardeau  \& Codis}{Uhlemann et~al.}{2022}]{Uhlemann:2022znd}
Uhlemann C.,  Friedrich O.,  Boyle A.,  Gough A.,  Barthelemy A.,  Bernardeau
  F.,   Codis S.,  2022, \mn@doi [Open J. Astrophys.]
  {10.21105/astro.2210.07819}, 6, 2023

\bibitem[\protect\citeauthoryear{Wechsler, Zentner, Bullock  \&
  Kravtsov}{Wechsler et~al.}{2006}]{Wechsler:2005gb}
Wechsler R.~H.,  Zentner A.~R.,  Bullock J.~S.,   Kravtsov A.~V.,  2006,
  \mn@doi [Astrophys. J.] {10.1086/507120}, 652, 71

\bibitem[\protect\citeauthoryear{White}{White}{1979}]{White:1979kp}
White S. D.~M.,  1979, Mon. Not. Roy. Astron. Soc., 186, 145

\bibitem[\protect\citeauthoryear{Yuan, Zamora  \& Abel}{Yuan
  et~al.}{2023}]{Yuan:2023llf}
Yuan S.,  Zamora A.,   Abel T.,  2023, \mn@doi [Mon. Not. Roy. Astron. Soc.]
  {10.1093/mnras/stad1275}, 522, 3935

\bibitem[\protect\citeauthoryear{Zeldovich}{Zeldovich}{1970}]{Zeldovich:1969sb}
Zeldovich Y.~B.,  1970, Astron. Astrophys., 5, 84

\bibitem[\protect\citeauthoryear{Zentner}{Zentner}{2007}]{Zentner:2006vw}
Zentner A.~R.,  2007, \mn@doi [Int. J. Mod. Phys. D]
  {10.1142/S0218271807010511}, 16, 763

\bibitem[\protect\citeauthoryear{de Lapparent, Geller  \& Huchra}{de~Lapparent
  et~al.}{1986}]{deLapparent:1985umo}
de Lapparent V.,  Geller M.~J.,   Huchra J.~P.,  1986, \mn@doi [Astrophys. J.
  Lett.] {10.1086/184625}, 302, L1

\makeatother
\end{thebibliography}

%\begin{thebibliography}{99}
%\end{thebibliography}

%%%%%%%%%%%%%%%%%%%%%%%%%%%%%%%%%%%%%%%%%%%%%%%%%%
%%%%%%%%%%%%%%%%% APPENDICES %%%%%%%%%%%%%%%%%%%%%

\appendix

\section{Effect of classification parameters on the letter function}\label{App_1}
\begin{figure}
\centering
\includegraphics[width=0.48\textwidth]{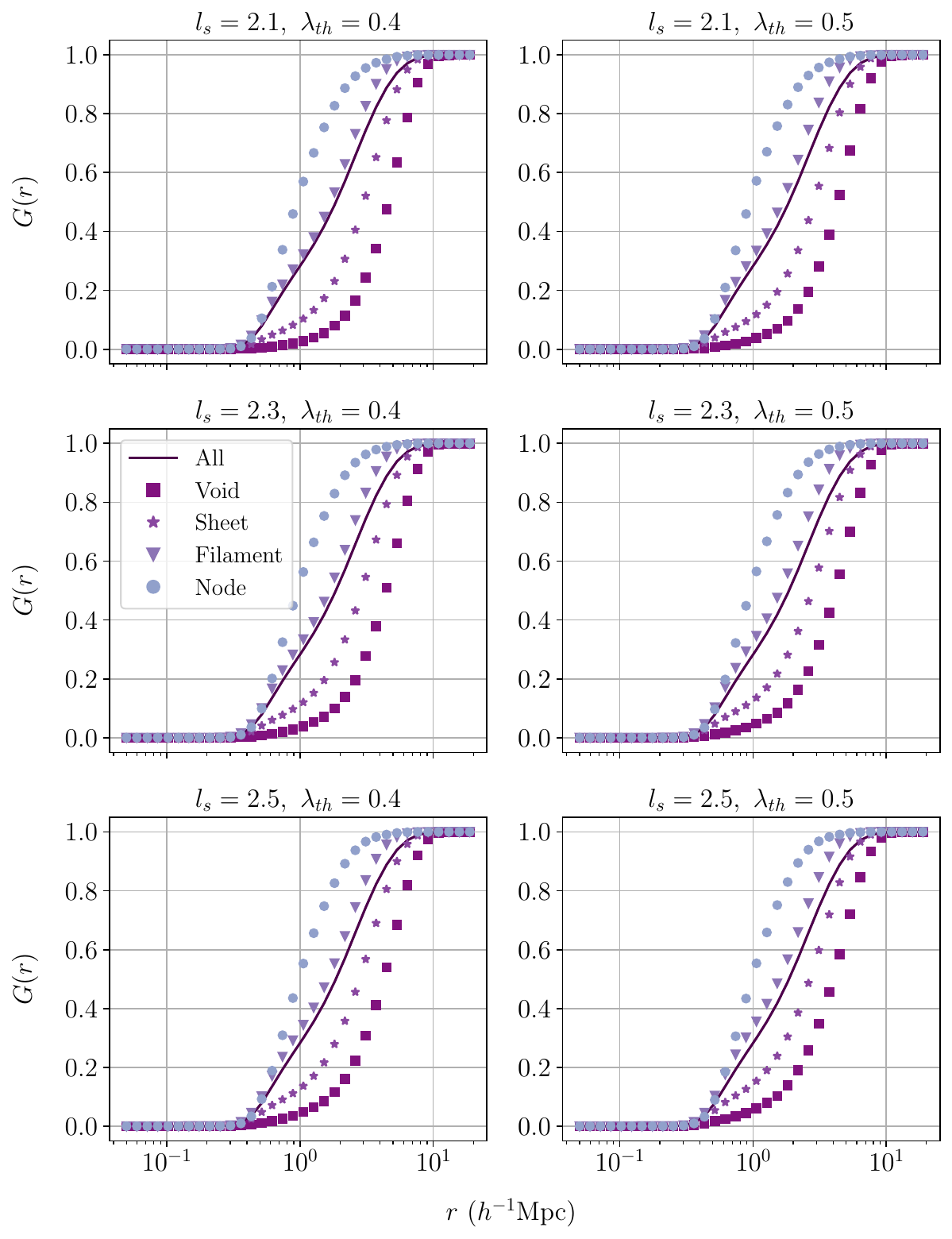}
\caption{The effect of classification parameters on $ G$ function in different cosmic environments for the $ \Lambda$CDM models. The values for the parameters are $ l_s = \ 2.1, \ 2.3, \  2.5 \ h^{-1} \rm{Mpc}$ and $ \lambda_{th} = \ 0.4, \ 0.5$. The solid lines stand for the whole halos. The squares, stars, triangles, and circles depict the $ G$ of the voids, sheets, filaments, and nodes, respectively.  The order and the shape of $ G$ functions stay the same, but a slight change in the values at different scales happens. This change is due to the change in the number density of halos in different environments. } 
\label{fig:App1}
\end{figure}

\begin{figure}
\centering
\includegraphics[width=0.48\textwidth]{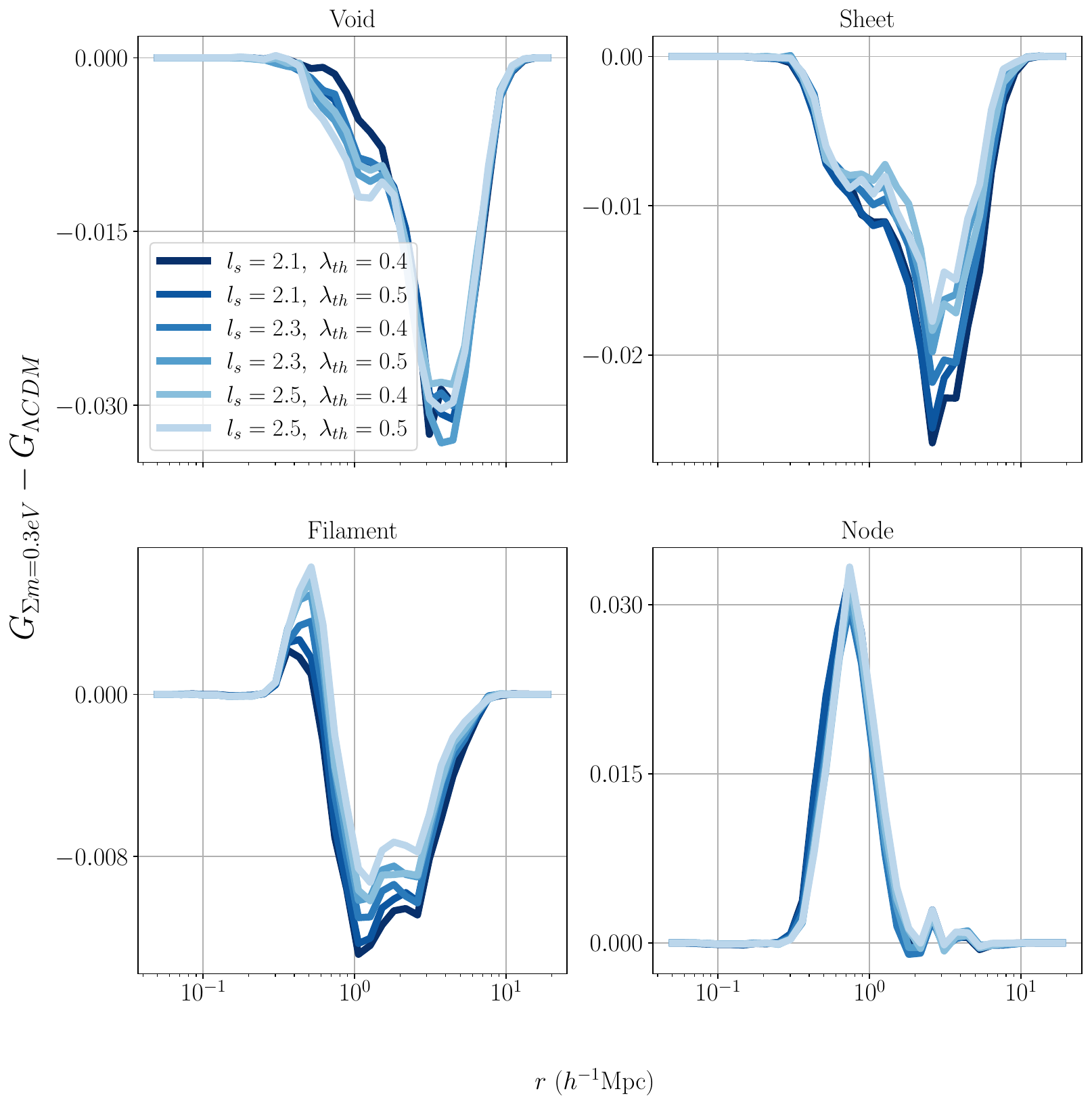}
\caption{The effect of classification parameters on the difference of $ G$ function between $ \Lambda$CDM model and $ \nu \Lambda$CDM model with total neutrino mass of $ 0.3$ eV in various cosmic environments. The shape again stays the same with a slight change in the values. This change is due to the change in the number density of halos in different environments. } 
\label{fig:App2}
\end{figure}

To check the impact of the classification parameters on the statistics we use in this work, we run the classification algorithm for six pairs of ($ l_s$, $ \lambda_{th}$), where $ l_s =0.4, 0.5 $  and $\lambda_{th} = 2.1, \ 2.3, \ 2.5 \  h^{-1} \rm{Mpc}$. All these values meet the criterion discussed in Section \ref{sec:CW-Classif}.  
\\
Figure \ref{fig:App1} shows the $ G$ functions in various environments for the $ \Lambda$CDM model. The markers of environments are the same as in Section \ref{Sec4}.  As evident in Figure \ref{fig:App1}, the shape and the hierarchy of $ G$ functions do not change.  A slight change in values of the $ G$ functions is detected, rooted in the variation of the number of halos in each environment.  
\\
Figure \ref{fig:App2} shows the impact of changing the $l_s$ and $ \lambda_{th}$ on the difference of the $ G$ functions between the standard model and the cosmologies with massive neutrinos. The order and the shape of the difference in $ G$ functions do not change, yet a slight change in the values of the difference is detected.  
\\
To conclude, if the parameters meet the criterion in Section \ref{sec:CW-Classif}., the results we represent in this paper do not change drastically. The letter functions computed in various environments could serve as a complementary probe for differing the extension of the standard model of cosmology.

%%****************FIG App-1**********************%%
%%****************FIG App-1**********************%%

%%****************FIG App-2**********************%%
%%****************FIG App-2**********************%%

%%%%%%%%%%%%%%%%%%%%%%%%%%%%%%%%%%%%%%%%%%%%%%%%%%
% Don't change these lines
\bsp	% typesetting comment
\label{lastpage}
\end{document}